# SWORD: Symmetry and Wyckoff-sequence of Ordered and Disordered crystals

Yuyao Huang[1], Wei Nong[1], Shuya Yamazaki[1], Martin Hoffmann Petersen[1], Jianghai Wang[1], Ruiming Zhu[1], Kedar Hippalgaonkar[1,2]*

[1] School of Materials Science and Engineering, Nanyang Technological University, 50 Nanyang Avenue, Singapore 639798, Singapore

[2] Institute of Materials Research and Engineering, Agency for Science, Technology and Research, Singapore 138634, Singapore

* Corresponding author: kedar@ntu.edu.sg

## Abstract

Novelty in materials discovery requires candidates to be distinct, non-redundant, and thermodynamically plausible. While crystallographic databases continue to expand in both size and complexity, making efficient and reliable novelty assessment has become increasingly difficult. This becomes particularly acute when crystallographic disorder is involved, as partial occupancies greatly enlarge the structure–composition space and obscure the identification of genuinely distinct structures. Here, we introduce SWORD, a symmetry-aware, Wyckoff-based string representation compatible with both ordered and disordered crystals. SWORD provides (i) standardization of symmetry-equivalent structural descriptions into a consistent label, (ii) explicitly represents co-occupying species on partially occupied sites, and (iii) quantifies complex disorder through a degree of mixing descriptor that captures continuous variation in site stoichiometry. These features enable efficient structure grouping, duplicate identification, and finer refinement of disordered structures. Benchmarking against existing fingerprint and structure-matching methods shows that SWORD remains invariant under identity-preserving transformations while retaining interpretable sensitivity to structural perturbations. In addition, SWORD shows competitive performance in associating unrelaxed and intermediate configurations with their final relaxed states along relaxation trajectories. This feature could enable more reliable novelty assessment directly from partially relaxed or even unrelaxed generated structures. Finally, SWORD was used to showcase its capability of disorder-aware database-scale deduplication and curation for the Inorganic Crystal Structure Database (ICSD). The curated ICSD would serve as the basis for the materials informatics and data-driven materials design in the era of artificial intelligence.

# 1. Introduction

The properties of crystalline materials, including electronic, ionic, and mechanical properties, are fundamentally governed by their structure. More generally, these properties are determined by a material's configuration within a high-dimensional structural–compositional space, where atomic arrangement and compositional variations together shape a material's behavior.[1] This raises the need to characterize similarity and novelty among crystalline materials, enabling meaningful comparison between related structures and reliable identification of genuinely distinct structures.

At the same time, decades of experimental characterization,[2,3] computational screening,[4,5] and recent structure-generation efforts[6-9] have driven the rapid expansion of crystallographic databases,[10] making it increasingly difficult to efficiently distinguish genuinely novel compounds from existing ones, an important requirement for claiming new materials discovery.[11] This can therefore lead to rediscovery of previously studied materials, limiting database diversity, and introduce bias into downstream data-driven analyses and machine-learning models.[12-14] These challenges have driven the development of crystal structure comparison methods, for which a useful representation should capture physically meaningful structural similarity while remaining invariant to equivalent descriptions of the same crystal. To achieve this goal, various crystallographic features have been explored to determine which are efficient and informative for representing and comparing crystalline materials.[15-17] One natural direction is to exploit crystallographic symmetry, where space groups and occupied Wyckoff positions provide a compact set of symmetry-distinct descriptors for compressing 3D atomic arrangements.[18-20] This aligns the encoding with crystallographic equivalence, so comparison focuses on essential structural information rather than arbitrary geometric representation. Another is graph-based information, where atomic connectivity and local environments offer a flexible way to encode both local and extended structural features.[21,22] From a methodological perspective, existing approaches can be reasonably grouped into two categories: direct structure-matching methods, which return numerical measures or thresholds to determine structural equivalence,[23,24] and fingerprint methods, which return explicit and indexable representations. Among the former, pioneering methods include STRUCTURE-TIDY,[25] COMPSTRU,[26] Pymatgen StructureMatcher[27], and AFLOW-XtalFinder.[28] These methods can resolve subtle differences between closely related crystals, but scale poorly when large numbers of structures must be compared across large databases. For large-scale deduplication and screening, however, such direct pairwise structure matching is not always necessary. More recent fingerprint methods, such as CLOUD[29], SLICES,[30] and graph-based descriptors including LeMat-BAWL,[22] which are based on simplified and compressed encodings that may sacrifice structural details needed to distinguish subtle differences, offer the efficiency and scalability required for large-scale applications while retaining meaningful structural accuracy.

Among various structural complexities in real crystals, a particularly challenging and widespread case is crystallographic disorder, which refers to deviations from a fully ordered periodic crystal structure.[31,32] Disorder is important as they often play a central role in modulating stability, enhancing transport, improving catalytic activity, reducing thermal conductivity, and in general enhance functional properties of materials.[33] Currently, approximately half of the entries in the inorganic crystal structure database (ICSD) exhibit some form of disorder.[34] Previous studies have shown that crystallographic disorder in inorganic crystals most commonly involves substitutional, vacancy, and positional disorder, or combinations of these.[35] In particular, positional disorder refers to cases where nearby sites intersect or are too close to be simultaneously occupied, while substitutional and vacancy disorder refer to cases where multiple atomic species and/or vacancies occupy the same crystallographic site, giving rise to partial occupancies. For brevity, the partially occupied sites are hereafter referred to as disordered sites. However, existing explicit representations generally do not natively encode partial occupancies, limiting their applicability to disordered structures. Although some structure-matching methods may have been extended to account for disorder,[9,24] they remain especially expensive as they often require enumerating more configurations and are still prone to matching failure.[9,36,37] As disorder greatly enlarges structural-compositional space of crystals, there is a pressing need for a crystal comparison method that encodes disorder naturally and scales to large datasets as well as generative models.

To overcome these limitations, we introduce Symmetry and Wyckoff-sequence of Ordered and Disordered crystals (SWORD), a string representation compatible with both ordered and disordered structures. By explicitly incorporating and quantifying crystallographic information under partial-occupancies, SWORD enables (i) compact and indexable representation of crystal structures, (ii) direct comparison and novelty assessment across crystals datasets, and (iii) scalable screening, classification, and cross-dataset retrieval across mixed ordered–disordered datasets. In the following, we first introduce the design principles and formulation of SWORD, then benchmark its robustness and distinctiveness under a range of identity-preserving transformations and structural perturbations against other structure-matching tools and crystal structure representations. Then, we also examine whether the structural identity captured by SWORD can be consistently related to physically meaningful relaxed structures. Finally, with the established performance of SWORD, we demonstrate its application to database-scale deduplication and curation in ICSD, one of the largest experimental crystal structure databases. Especially useful is using it to ascertain the Degree of Mixing (DOM) that allows distinguishing different materials with the same SWORD label but differing in site occupancy (useful, for example in, doped materials and solid solutions).

## 2. Methods

### 2.1 SWORD representation

Infinite periodic crystal structures are naturally described by space groups and occupied Wyckoff positions. Recent crystal generative models have likewise adopted symmetry-based representations built on this crystallographic principle.[7,8] In the same spirit, SWORD leverages this information to identify symmetrically equivalent structures, as visualized in Fig. 1. AFLOW Prototype similarly represents structures using the space-group number together with Wyckoff-position information and composition.[38] The key distinction, however, is that AFLOW prototype uses anonymous formula (no elements) that does not retain the correspondence between Wyckoff sites and the atomic species occupying them. In contrast, SWORD adopts a Wyckoff-site-resolved, element-wise sequence representation, in which independent occupied Wyckoff positions are traversed in sequence and their elemental occupations are explicitly assigned to each site.

As shown in Fig 1a, in the SWORD string, each Wyckoff position is directly paired with its occupied element. For example, disordered $Li_{1.34}Mn_{0.66}O_2$ (ICSD collection code 149600) is written as f4_d_e3_15_{3O,Li}_Li_{2(Li+Mn),Mn}. Here, f4_d_e3 and 15 specify the occupied Wyckoff-site sequence and space-group number, with Wyckoff position f occupied by three O sites and one Li site, Wyckoff position d occupied by Li, while Wyckoff position e contains two compositionally disordered Li/Mn sites and one fully occupied Mn site. In this way, SWORD provides a unified representation for both ordered and disordered crystals. Structures that share the same disordered-site pattern but differ in site stoichiometries at disordered sites are further distinguished by the metric introduced in the following section. It should be noted that SWORD is designed for symmetrized unit-cell descriptions rather than low-symmetry supercell representations.

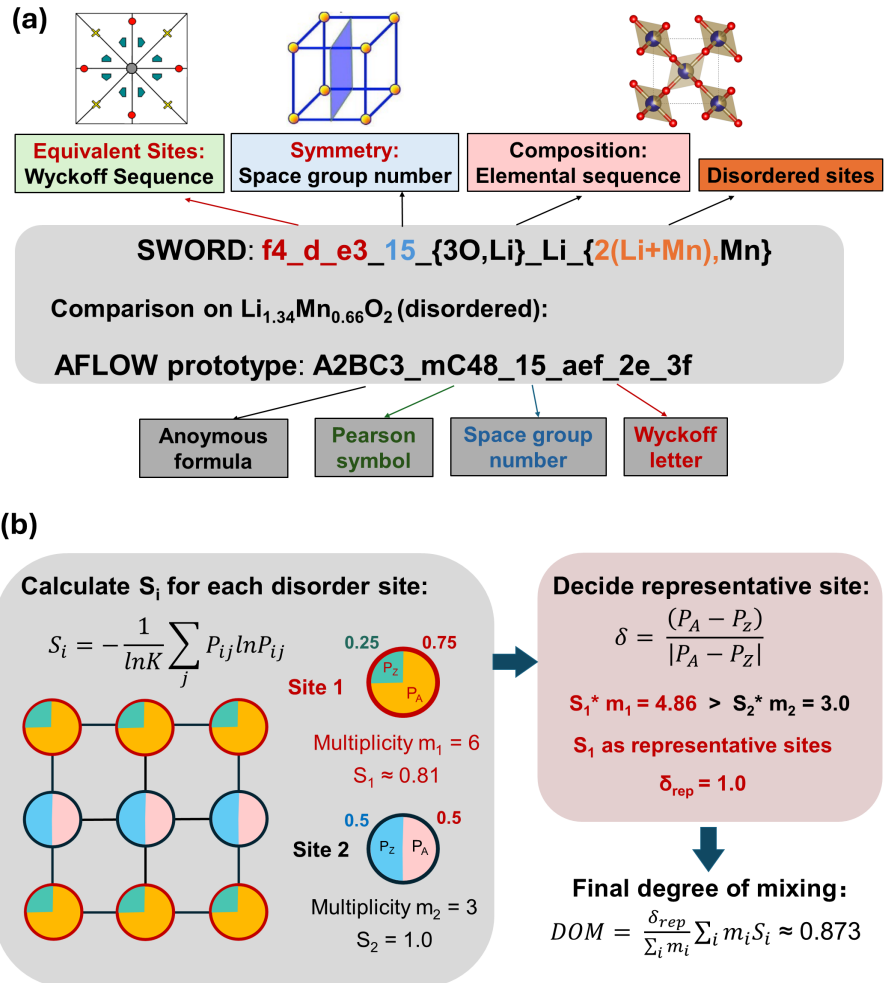

**Figure 1** (a) Comparison of SWORD and AFLOW prototype for disordered $Li_{1.34}Mn_{0.66}O_2$. SWORD encodes a crystal as a site-resolved sequence of Wyckoff–element assignments with explicit disordered-site occupations, whereas the AFLOW prototype provides an anonymous symmetry-based description without preserving site-resolved chemistry. (b) Workflow for calculating the degree of mixing (DOM). DOM is applied to distinguish disordered structures that share the same SWORD label but differ in site occupancies. Structure-level mixing scores across all sites are combined using site mixing and corresponding multiplicities, and representative-site bias to obtain the final DOM value.

## 2.2 Degree of mixing

To distinguish disordered structure templates, the SWORD label serves as the unique identifier for novelty assessment. However, even within the same SWORD label, multiple entries may exist with different site occupancy ratios, which do not necessarily constitute meaningful novelty. To differentiate such cases, we define a degree of mixing (DOM) for a disordered structure, based on normalized Shannon entropies (Pielou's evenness)[39] across all partially occupied Wyckoff sites. Although our method can identify and label structures with positional disorder, such cases are not further considered here, as their inclusion would overestimate the Shannon entropy. As shown in Fig. 1b, this quantity characterizes how evenly different species occupy each disordered site. DOM is not intended for direct comparison of disordered structures across different SWORD labels, because the underlying symmetry framework and disordered sites are not directly commensurate. For a partially occupied Wyckoff site $i$, its normalized Shannon entropy $S_i$ can be calculated as:

$$S_i = -\frac{1}{lnK}\sum_j P_{ij}lnP_{ij}$$

where $K$ is the number of species on site $i$, $P_{ij}$ is the partial occupancy of species $j$ on the site $i$. The normalization coefficient $lnK$, derived from maximized Shannon entropy under perfect mixing ratio, places sites with different numbers of co-occupying species on the same scale. In Fig. 1b, a hypothetical configuration containing two disordered sites with multiplicities of 6 and 3 yields $S_1 \approx 0.81$ and $S_2 = 1.0$, respectively, showing that site 2 is more evenly mixed.

The DOM can be calculated as the multiplicity-weighted average of site Shannon entropies across all disordered sites in the structure:

$$DOM = \frac{\delta_{rep}}{\sum_i m_i}\sum_i m_i S_i$$

where $m$ is the multiplicity of Wyckoff site $i$, and $\delta_{rep}$ is the mixing bias indicator, explained later. This weighting reflects the fact that disorder on higher-multiplicity sites contributes more strongly to the overall structure-levelal mixing. In the example of Fig. 1b, site 1 is selected as the representative site because $m_1 S_1 = 4.86$ is greater than $m_2 S_2 = 3.0$.

Because Shannon-entropy based DOM is a scalar measure of the magnitude of mixing, different structures may still share the same DOM value even when the mixing is biased toward different species. To retain this bias information, a mixing bias indicator $\delta_{rep}$ is defined as:

$$\delta_{rep} = \frac{(P_A - P_z)}{|P_A - P_Z|}$$

where $P_A$ and $P_Z$ are the partial occupancies of two species A and Z with the most extreme occupancies on the representative disordered site. The sign direction is assigned according to alphabetical ordering of the species labels, with $P_A$ corresponding to the species with the alphabetically smaller symbol. The choice of alphabetical ordering is not special for anything but simply for consistency. Notably, $\delta_{rep}$ is assigned as 1 when variables are equal to each other ($P_A = P_Z$).

## 2.3 Identification of equivalent site and disordered types

General-purpose crystallographic toolkits such as Spglib[40], and Pymatgen[27] provide the basis for analyzing symmetry and structure standardization. In SWORD, Pymatgen is used to identify symmetry and convert the input structure to a symmetrized conventional-cell description. Although Pymatgen is able to parse disordered structures, it does not directly provide a high-level interface for conveniently extracting and processing all the information required for disorder analysis in SWORD. This limitation becomes more serious when some structures contain inconsistent coordinates for the disordered site, numerical noise, or excessively close atomic sites, which may obscure crystallographic equivalence and potentially lead to incompletely or incorrectly grouped disordered-site assignments. Therefore, SWORD identifies equivalent sites directly from the symmetrized crystallographic coordinates before subsequent disorder encoding. During this process, the presence of disorder is also determined, and disordered sites are further classified as substitutional, vacancy, or positional disorder.

For identification of disordered type, SWORD adopts different tolerance criteria. More details are given in S2 of Supporting Information. For substitutional disorder, co-occupying species are identified when the Euclidean distance between candidate sites falls within a predefined tolerance. For vacancy disorder, a specific vacancy threshold is applied to sites with total occupancies below 1.0. When the vacancy fraction exceeds this threshold, vacancy is explicitly encoded in the SWORD label as 'VAC'. Positional disorder is identified when the candidate sites satisfy a case-specific distance criterion defined from the separation between sites and effective radii of the involved species, consistent with the framework of Antypov et al.[35]

## 2.4 Standardization of Wyckoff sequence

A typical challenge in symmetry-based structure representations arises from the non-uniqueness of the conventional crystallographic description. Alternative conventional-cell settings due to crystal axis permutation (non-standard cell settings), and various origin choices, equivalent translation of Wyckoff sites under Euclidean or affine normalizer of the space group can cause symmetry-equivalent structures to be described in different Wyckoff-letter sequences.[41] Therefore, direct comparison of raw Wyckoff letters is insufficient. Among the possible equivalent descriptions of the same space group, Hall settings provide an explicit way to distinguish alternative symmetry settings.[42] To ensure that symmetry-equivalent descriptions yield the same SWORD label, we standardize the Wyckoff description by mapping symmetry-equivalent Wyckoff decorations into a unique representation before assembling the final SWORD string.

The standardization is carried out in two stages. In the first stage, for a symmetrized structure with identified space group, the affine transformation matrices connecting alternative Hall settings are applied to lattice basis and atomic coordinates. After each transformation, Hall-equivalent candidates — the set of reassigned Wyckoff sequences — will be generated. Then, these candidates are mapped to the default Hall reference, yielding a Hall-normalized representation. In the second stage, starting from this Hall-normalized representation, the affine transformations associated with transformed Wyckoff-position descriptions of the same space group are considered. These transformations generate the remaining symmetry-equivalent relabeling of the Wyckoff sequence that are not removed by the Hall normalization alone. The resulting candidates are then reduced to a single standardized Wyckoff sequence, from which the final SWORD label is assembled. Details of the Hall-settings, affine transformations and standardization procedure are provided in S3 of Supporting Information.

## 3. Results

### 3.1 Benchmarking the performance of SWORD

With SWORD formalized, we systematically evaluate its performance following the benchmarking scheme adopted in a previous study.[22] A useful crystal fingerprint should remain invariant under moderate perturbations and identity-preserving transformations, while retaining sufficient sensitivity to distinguish structures that become genuinely different under severe perturbations. The benchmark is therefore structured around three complementary perspectives, all evaluated on ordered crystal structures: robustness under controlled perturbations and transformations, scalability as dataset size increases, and the ability to identify the same structure across different structure relaxation states. The structure relaxation here refers to the energy and force minimization process for a given original or perturbed structure, as defined in Section1 of the Supporting Information.

The benchmarked methods include BAWL, short-BAWL, Pymatgen StructureMatcher, PDD, SLICES, and SWORD. For testing robustness, four categories of perturbations were considered: i) rigid site translation, ii) isotropic lattice strain, iii) symmetry operations, and iv) Gaussian random noise applied to atomic coordinates and lattice vectors. These perturbation methods are illustrated in Fig. 2a. For each method and each perturbation level, a set of perturbed structures is generated from the original structure and then compared with the original one. The first three probe invariance under identity-preserving transformations, whereas the Gaussian-noise tests examine the sensitivity of each method to progressively stronger structural perturbations. The matching rate at a given perturbation level is defined as the proportion of perturbed structures that remain matched to the pristine structure, and the reported curve is obtained by averaging this quantity over all sampled structures. A higher match rate and slower decline indicate greater robustness under the tested perturbations. For this benchmark, 200 structures were randomly sampled from LeMat-Bulk, a dataset comprised of 6 million structures from the Materials Project, Alexandria and OQMD.[10,43,44] For each sampled original structure, perturbations were applied at different noise levels, and 10 independent perturbed structures were generated randomly for each noise. Reported results are averaged over these 10 perturbed structures, and shaded bands indicate the corresponding lower and upper bounds.

As shown in Fig. 2b, under rigid site translation, SWORD retains a match rate of 1.0 across the full tested range, which is the expected and desirable behavior for an invariant crystal fingerprint, matching the performance of BAWL, Short-BAWL, and Pymatgen StructureMatcher. By contrast, SLICES and PDD show clear losses in consistency. For isotropic strain, SWORD, BAWL, Short-BAWL, and Pymatgen StructureMatcher all retain 1.0 match rate across the tested range, whereas SLICES shows a modest but systematic loss of consistency, and PDD exhibits the strongest breakdown. This result indicates that SWORD is insensitive to the absolute origin of the atomic coordinates and remains fully invariant across the tested isotropic strain range, consistent with the design of SWORD, as lattice information is not explicitly encoded and isotropic scaling does not alter the underlying symmetry of the structure. For symmetry operations, SWORD, BAWL, Short-BAWL, and Pymatgen StructureMatcher all achieved a match rate of 1.0, whereas PDD achieved only partial invariance with a match rate of 0.8 and SLICES showed a pronounced loss of invariance. These benchmark results show that, as a symmetry-aware representation, SWORD preserves invariance as reliably as other well-performing methods, which is essential for both crystal fingerprinting and deduplication.

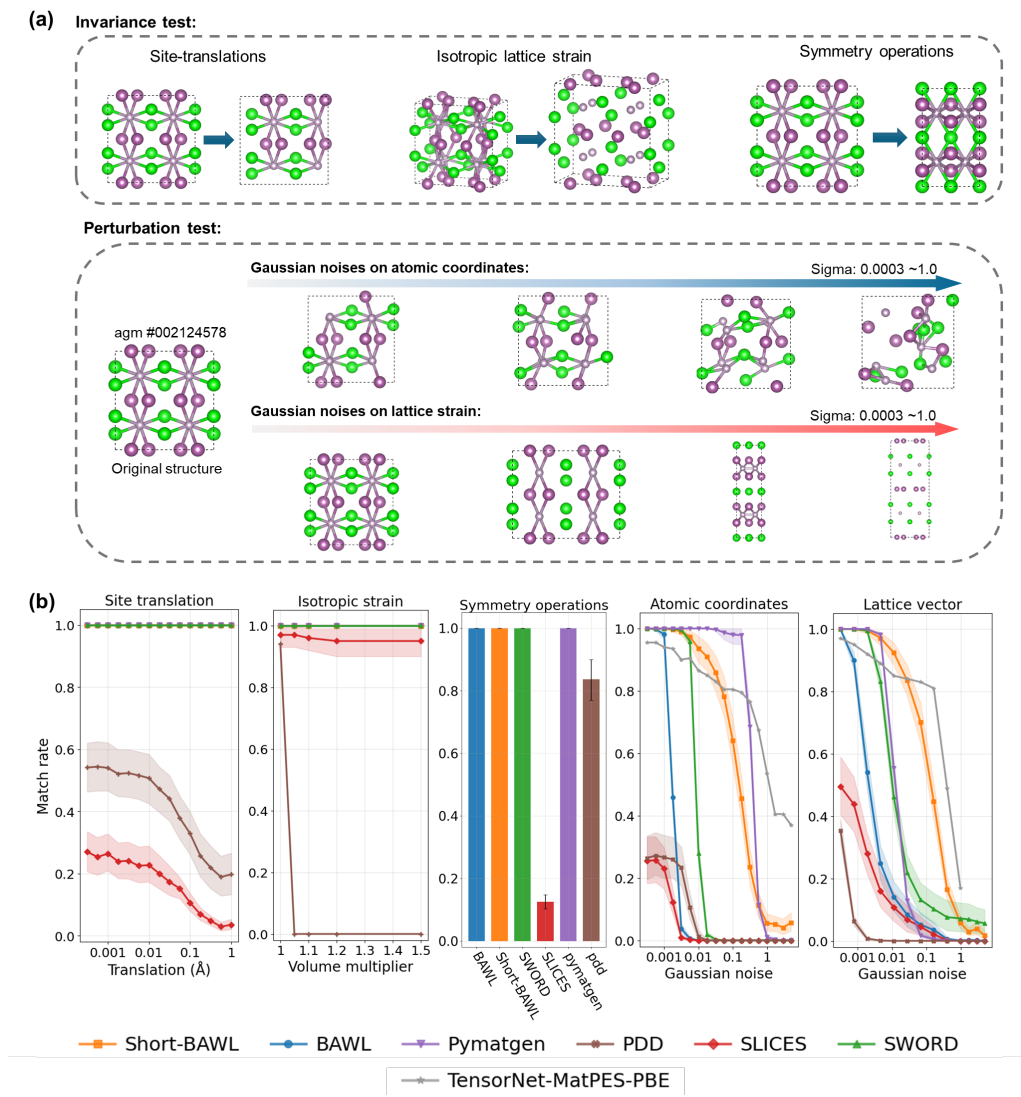

**Figure 2** (a) Schematic illustration of the invariance and perturbation tests, including site translations, isotropic lattice strain, symmetry operations, and Gaussian noise applied to atomic coordinates or lattice vectors. (b) Match rates of different methods under each transformation or perturbation condition. At each noise level, match rates are averaged over 10 batches of 200 perturbed samples. Shaded bands around curves denote lower and upper bounds across batches.

As shown in Fig. 2b, for Gaussian noise applied to atomic coordinates, Pymatgen StructureMatcher maintains the highest match rate overall across the tested perturbation range, while Short-BAWL remains the most robust among the fingerprint methods. Under the present benchmark setting, SWORD remains more stable than BAWL, SLICES, and PDD. A similar trend is observed when Gaussian noise is applied to lattice vectors. For symmetry-dependent representations, such as SWORD and BAWL, both the curves show a clear cutoff because their response to random perturbations is influenced by the tolerance used for symmetry detection in Pymatgen; further details are provided in Section S1 of the Supporting Information. Usually, a more stringent tolerance would lead to space-group change and label mismatch even at lower noise level. Compared with graph- or geometry-based methods such as Short-BAWL and Pymatgen StructureMatcher, SWORD is more sensitive to perturbations in the default symmetry tolerance. This sensitivity, however, can be tuned to suit different applications or material datasets. This feature can be advantageous when finer structural discrimination is needed, particularly for distinguishing closely related structures that share similar geometry but differ in symmetry or site ordering. The benchmark results highlight that SWORD as a symmetry-aware representation with proper standardization of Wyckoff descriptions remains robust under identity-preserving transformations while retaining interpretable sensitivity to random perturbations. Given an appropriate symmetry tolerance, symmetry-aware representations can serve as reliable fingerprints for large-scale duplicate detection and structural comparison.

In addition to direct label matching, a relaxation-based reference is introduced to provide a complementary view of structural equivalence under random perturbations. In this reference, the perturbed structures are relaxed by TensorNet-MatPES-PBE machine learning interatomic potentials (MLIP) and then compared with their corresponding relaxed original structures (see S1 of the supporting information). A structure is considered as mismatched if the relaxation energy difference exceeds 0.15 eV/atom, which lies within empirical metastability window adopted in some previous studies,[45] or if the relaxed structure is no longer equivalent to the original according to StructureMatcher. This setup effectively probes whether a perturbed structure remains in the same basin of the potential-energy surface, such that relaxation can recover the original relaxed configuration. After relaxation, trajectories from calculations whose final structures remain matched to the relaxed original structures according to the two criteria are referred to as "valid trajectories" and will be used for subsequent analysis. As highlighted in Fig. 2b, the corresponding grey curve decreases more gradually than those based on direct matching, for both atomic-coordinate noise and lattice perturbations. This indicates that some perturbed structures, although substantially displaced from the potential energy minimum, can still relax back to the same structural basin. These results indicate that exact label invariance is generally stricter than equivalence defined through relaxation. However, this baseline serves only as an auxiliary guide rather than a definitive ground truth, since it depends on the choice of relaxation method and equivalence criteria.

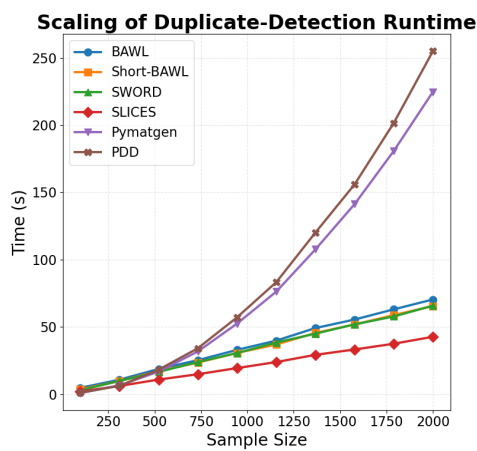

**Figure 3** Runtime scaling of duplicate-detection methods on a single CPU core as a function of sample size.

Fig. 3 reports the runtime on a single CPU core required by each method for processing and aggregating structures as a function of the sample size. Fingerprint-based methods exhibit an approximately linear scaling trend and remain substantially faster at larger batch sizes than pairwise matching methods that scale quadratically. Among the fingerprint methods, SWORD shows runtime comparable to BAWL and Short-BAWL. Overall, these results indicate that SWORD combines competitive effectiveness with favorable scalability for database-scale duplicate identification.

## 3.2 Structural identity and uniqueness

As SWORD is built on the Wyckoff representation, it compresses structure information through space-group symmetry, and some information loss is therefore inevitable. This issue is well known for symmetry-based representation and implies that the relevant objective is not strict invertibility for arbitrary geometries, but a robust correspondence between the representation and the energy-minimized state.

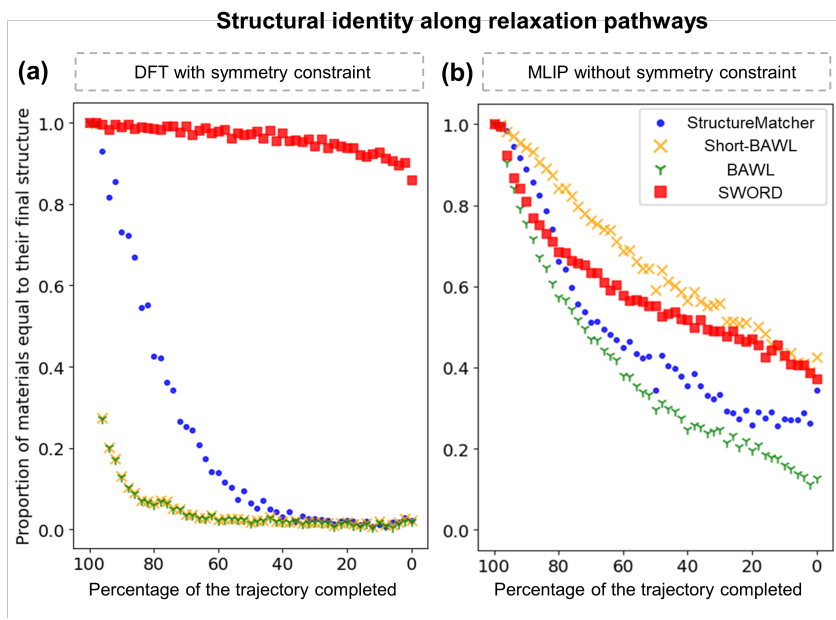

**Figure 4** Comparison among different fingerprints and structure matcher methods for the proportion of structures matched to their final relaxed state as a function of relaxation progress on (a) DFT symmetry-constrained and (b) symmetry-unconstrained MLIP relaxation trajectories.

Based on this assumption, we perform a complementary analysis for fingerprint and structure-matching methods that had shown satisfactory performance in previous benchmark, including BAWL, Short-BAWL, Pymatgen StructureMatcher, and SWORD. We evaluate this property by testing whether SWORD and other methods can match un-relaxed or intermediate configurations to the corresponding final relaxed structure. Because relaxation pathways may practically depend on the implementations of energy prediction (e.g., by DFT or MLIP) and whether symmetry is externally constrained during relaxation, two test datasets were constructed: one consisting of 1000 density functional theory (DFT) symmetry-constrained relaxation trajectories from MPtraj database[46] and the other consisting of 1000 symmetry-unconstrained MLIP trajectories retrieved from perturbed benchmark structures generated in this study. The unconstrained MLIP relaxations were performed as an alternative to unconstrained DFT relaxations for accelerated benchmarking. It should be noted that symmetry constraint is imposed by default during DFT relaxations but is not during MLIP relaxations. The match rate for each method along the relaxation progress was then computed only over the samples in the valid trajectories as obtained in the previous relaxation-based reference.

As shown in Fig. 4a, in the DFT relaxed trajectories, SWORD maintains consistently high match rate across the whole relaxation progress, whereas StructureMatcher, Short-BAWL, and BAWL remain at a low rate and StructureMatcher achieves comparable match rate only after when the end of relaxation progress approaches. Symmetry-based SWORD might have gained benefits from symmetry constraints. One can see that the match rate of SWORD does drop slightly in the case of symmetry-unconstrained relaxations, as shown in Fig. 4b. Notably, the robustness of SWORD still holds owing to its much higher match rate than most of the others. In the MLIP relaxed trajectories, SWORD remains comparable to Short-BAWL during the initial stages of the relaxation trajectory, indicating that it can represent energy-minimized states reasonably well even for configurations that remain far from convergence, despite the absence of symmetry constraints during optimization.

These results indicate that SWORD captures relaxation-equivalent structural identity and is more effective at associating intermediate states with their final relaxed basin. This is particularly relevant in the context of crystal structure generation, where initial outputs are often unrelaxed and therefore displaced from their energy-minimized states. Given the computational cost of full relaxation, it is often advantageous, when feasible, to perform novelty checks against fully converged reference structures using unrelaxed outputs. A higher match rate to the corresponding converged states at relatively early ionic steps suggests that reliable novelty assessment can already be achieved from partially relaxed or even unrelaxed structures. This, in turn, indicates that SWORD enables more robust novelty checks on unrelaxed generated structures compared to other methods.

### 3.3 Deduplication and curation of ICSD

The practical value of SWORD should be evaluated not only through specific benchmark tasks, but also from its performance in downstream database-scale tasks. Its disorder-aware encoding enables large-scale deduplication and curation of ICSD, a task not readily supported by most existing fingerprint methods. As an experimental crystallographic database, the ICSD contains repeated or closely related entries in both ordered and disordered forms. Because the ICSD has accumulated data from diverse literature sources, the entries exhibit substantial variation in crystallographic description and refinement quality. These characteristics make ICSD curation a demanding yet particularly relevant testbed for evaluating SWORD. In the following, we evaluate SWORD as a disorder-aware framework for standardizing, grouping, and curating ICSD entries at database scale.

As illustrated in Fig. 5a, the workflow of processing ICSD can be divided into three stages. In stage I, 241,152 raw ICSD entries are lightly preprocessed to exclude nonstandard formatting records and remove entries containing hydrogen, placeholder species, or isotopic elements, resulting in 214,637 entries. In stage II, the recorded crystallographic information files (CIFs) are standardized and return SWORD labels together with related information. Because the ICSD provides experimentally assigned symmetry information, this stage operates directly on the raw CIFs rather than through Pymatgen interface. The workflow also identifies substitutional, vacancy, positional disordered type and compute DOM for subsequent comparison. In stage III, the labeled entries are post-processed for curation and deduplication. Entries that contain inconsistent disordered sites, non-physical partial occupancies, too-close atoms and large numerical noise are collected as anomalous records for further inspection, amounting to 3,661 entries. The categories and detailed descriptions of the anomalous record types can be found in S3 of the Supporting Information. In total, SWORD partitions the processed ICSD dataset into 102,178 identical-label groups, as defined in S1 of supporting Information. At this stage, groups containing more than one entry are regarded as duplicate groups at the SWORD-label level. For disordered entries, however, sharing the same SWORD label does not necessarily imply full redundancy, because entries within the same label group may still differ in stoichiometry and occupancy distribution. Therefore, disordered entries within each identical-label group are further refined using DOM, and in the present work, only entries with identical DOM values are collapsed into a single representative. After this additional refinement, the final curated dataset contains 127,056 entries.

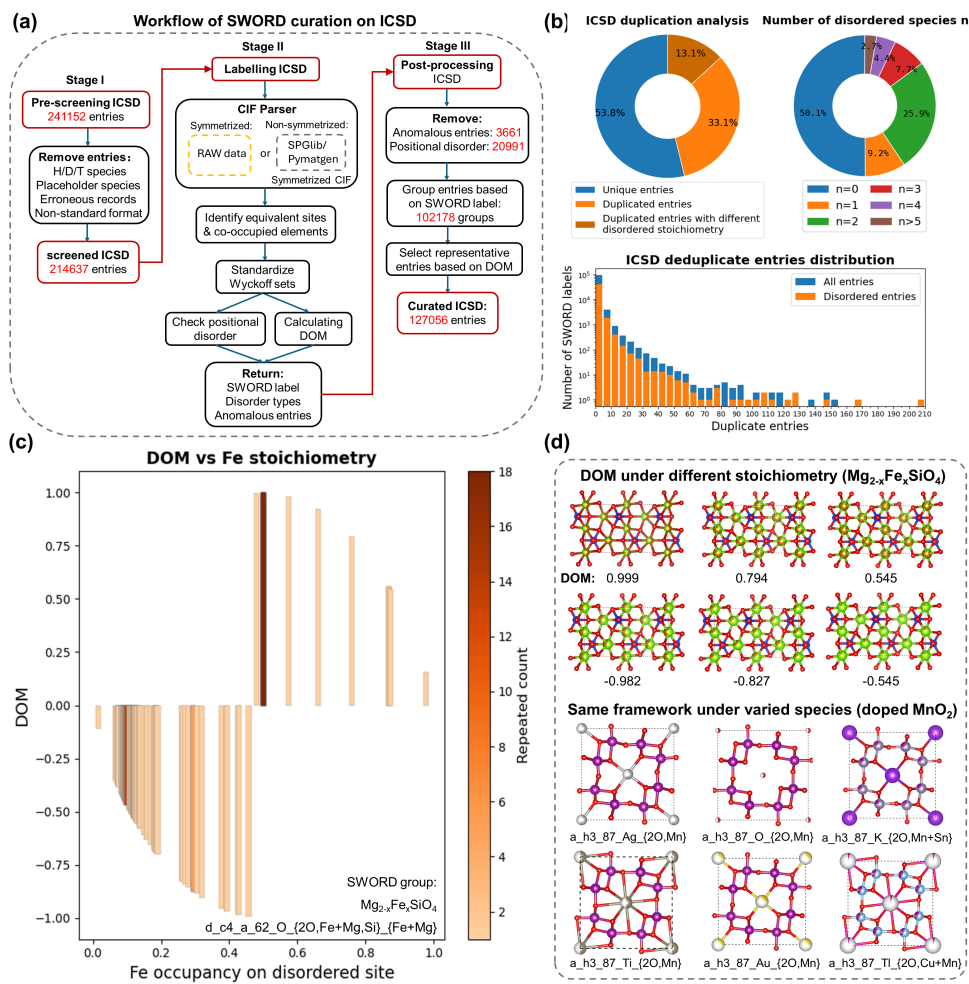

**Figure 5** (a) Workflow of SWORD-based ICSD deduplication and curation. (b) Statistics of ICSD duplication results, number of disordered species, and distribution of duplicate-group sizes (For clarity, the largest duplicate-group size 290 is not shown on the x axis). (c) DOM as the function of Fe stoichiometric variation illustrated in the case of $Mg_{2-x}Fe_xSiO_4$ within the same SWORD framework and (d) Examples of $Mg_{2-x}Fe_xSiO_4$ under different DOM, and metal-doped $MnO_2$ with identical frameworks but different disordered species. Vacancy components are omitted from the displayed labels for clarity limitation.

At the database scale, SWORD reveals a highly uneven duplicate landscape in the ICSD. As shown in Fig. 5b, distinct SWORD labels account for 53.8% of the processed ICSD entries, while the remaining 46.2% correspond to duplicated instances of labels already represented elsewhere in the dataset. Among the processed ICSD entries, 50.1% are classified as ordered and 49.9% as disordered, consistent with previous research.[35] Within the disordered subset, two disordered species is the most common case, accounting for 25.9% of all entries, while structures with three or more disordered species become less frequent. The remaining 9.2% represent pure vacancy-type disorder with only one disordered species. In addition, 84% of SWORD labels contain only 2-5 duplicates. Although the number of groups decreases rapidly as the number of duplicates increases, the largest ordered and disordered duplicate groups still contain 155 and 290 entries, respectively. To illustrate this, we examine the material group $Mg_{2-x}Fe_xSiO_4$ in Fig. 5c with 145 duplicates and corresponding DOM within the group, where the DOM is plotted as a function of the Fe occupancy at the Fe/Mg disordered site. This example shows that structures sharing the same SWORD label may still span a broad range of site occupancies. For binary disorder, the signed DOM provides a clear separation between opposite occupancy biases. The upper panel of Fig. 5d also clearly illustrate this directional effect in the $Mg_{2-x}Fe_xSiO_4$ examples, where the corresponding Fe/Mg occupancy proportions are reflected in both the sign and magnitude of the DOM values. For ternary or higher-order disorder, the scalar DOM still quantifies the overall degree of mixing, while the sign is assigned from the representative disordered site with the largest weighted contribution to the overall DOM. Therefore, DOM can also be used to set-up user-controlled intervals, allowing different levels of refinement when grouping similar but complex disordered states.

At the same time, SWORD preserves distinctions between entries that share the same or closely related Wyckoff framework but differ in the chemical species of the disordered sites, so that symmetry-level similarity does not obscure site-resolved chemical differences, as illustrated by various metal-doped $MnO_2$ structures in lower panel of Fig. 5d. These results indicate that SWORD provides a practical grouping key for large-scale database analysis while retaining both prototype-level grouping and chemically meaningful variation for deduplication and curation. Moreover, the curated, deduplicated ICSD by SWORD can be used to identify and evaluate novelty of ordered and disordered crystal structures produced from different crystal structure databases and crystal generative models.

## Conclusion

In this work, SWORD (Symmetry and Wyckoff-sequence of Ordered and Disordered crystals) is introduced as a symmetry-aware, Wyckoff-based string representation compatible with both ordered and disordered crystals. By combining standardized Wyckoff-sequence encoding with explicit representation of disordered sites, SWORD maps symmetry-equivalent crystallographic descriptions to a unified label. It also converts multi-element and multi-site disorder into a quantifiable and comparable descriptor, DOM, enabling further refinement of site-stoichiometric variation under identical-label disordered structures. Benchmarking against existing fingerprint and structure-matching methods showed that SWORD remains fully invariant under identity-preserving transformations, maintains a practical balance of robustness and structural discrimination under random perturbations, and exhibits favorable computational scaling for database-scale applications. In addition, relaxation-trajectory analyses indicate that SWORD can more consistently associate intermediate or perturbed structures with their final relaxed state, supporting its ability to capture physically meaningful structural identity beyond exact geometric matching.

These results demonstrate the practical value of SWORD for large-scale deduplication and curation of ordered and disordered structures, particularly for some dataset curation that are not readily supported by existing methods. For crystallographic dataset, SWORD provides a unified disorder-aware workflow that combines standardized labeling, disorder-type identification, duplicate grouping, and DOM-based refinement of disordered duplicates. Applied to ICSD, SWORD shows that 46.2% of the processed entries correspond to duplicates at the SWORD-label level, while disordered duplicates can be further refined through user-controlled DOM intervals to separate entries with different site stoichiometries. Overall, SWORD provides a practical disorder-aware framework for duplicate identification and novelty assessment between experimental, theoretical, and generated crystal datasets. More broadly, its symmetry-based formulation also provides a natural foundation for future extensions toward analysis of ordered and disordered phases related through space-group hierarchy, including the identification of possible disordered parent structures from ordered structures.

## Data Availability

The crystal structure data used in this study were obtained from the Inorganic Crystal Structure Database (ICSD), version 2025.2_v5.5.0. ICSD entries are available at https://icsd.products.fiz-karlsruhe.de under a commercial license.

## Code Availability

The implementation of SWORD, together with the ICSD data curation pipeline used in this work, is available at https://github.com/YuyaoHuang330/SWORD.

## Acknowledgement

K.H. acknowledges funding from the MAT-GDT Program at A*STAR via the AME Programmatic Fund by the Agency for Science, Technology and Research under Grant No. M24N4b0034. This research is supported by the National Research Foundation, Singapore under its AI Singapore Programme (AISG Award No: AISG3-RP-2022-028).

# Supporting Information

## S1. Glossary

**- Identity-preserving transformations** - Transformations that are treated in this benchmark as preserving structural identity. In this work, these include rigid site translation, isotropic lattice strain, and symmetry operations.

**- Symmetry tolerance** - The tolerance parameter used in symmetry detection by Pymatgen's SpacegroupAnalyzer. In the present benchmark, SWORD uses symmetry tolerance of 0.05. The default value is 0.01 in Pymagtgen.

**- Original structure** - Structure directly sampled from LeMat-Bulk without any transformation and perturbation

**- Perturbed structure** - A structure obtained from an original structure after applying a defined operation, including identity-preserving transformations and random-noise perturbations.

**- Sensitivity -** The extent to which a method responds to identify preserving transformation or perturbations. In this work, higher sensitivity means that matching is lost at lower perturbation levels, whereas lower sensitivity indicates greater tolerance/robustness to perturbations.

**- Identical-label group** - A group containing one or more entries that share exactly the same SWORD label.

**- Duplicate group** - At the SWORD-label level, an identical-label group containing more than one entry.

## S2. Identification of equivalent sites and classification criteria of disordered types

A Wyckoff site is a symmetry-unique generator whose full atomic positions can be recovered by applying space-group operations. The identification of disordered sites in SWORD is built on an orbit-based framework, in which crystallographic equivalence is determined from symmetry-generated atomic orbits rather than from direct pair-wise comparison of Wyckoff-site coordinates. This design ensures that equivalent sites are recognized consistently even when their fractional coordinates appear different due to non-standard formatting, lattice boundary condition or numerical noise, which are known to obscure equivalence in raw crystallographic representations.

First, candidate Wyckoff sites are grouped by same Wyckoff letters. For each candidate coordinate r, the symmetry-equivalent orbits $O(r)$ are generated from the full set of symmetry operations under the corresponding space group:

$$O(r) = \{\ gr\ |\ g \in G\ \}$$

where G denotes the space group. The resulting orbit is then reduced under periodic boundary conditions to remove duplicates arising from numerical noise. This reduction is performed using a fractional-coordinate tolerance of $10^{-4}$.

To determine whether two candidate Wyckoff sites $i$ and $j$ correspond to the same disordered site, their orbits are compared as sets of fractional coordinates. Specifically, we define their set-to-set difference in periodic fractional-coordinate space as:

$$D(O_i, O_j) = \max_{r \in O_i} \min_{r' \in O_j} \| r - r' \|_{PBC}$$

where $\|\cdot\|_{PBC}$ denotes the shortest Euclidean distance between two fractional coordinates under lattice periodicity. For each point r in orbit $O_i$, the inner minimum selects its nearest point in orbit $O_j$, and the outer maximum then takes the largest of these nearest-point distances. If the two orbits difference $D(O_i, O_j)$ within a default site tolerance of $10^{-4}$, they are merged into the same equivalent site.

## S2.2 Identification of substitutional and vacancy disorder

Once equivalent sites are grouped, substitutional and vacancy disorder are identified directly based on the grouped orbit. If multiple non-vacancy species share the same equivalent site, that site is classified as substitutionally disordered. For each equivalent-site group, the total occupancy can be easily obtained by summing the occupancies of all species assigned to that site. If the remaining unoccupied fraction, defined as 1.0 minus the total occupancy of that equivalent-site group, exceeds the vacancy threshold, the site is classified as vacancy-disordered and the vacancy is explicitly encoded into the SWORD label as 'VAC'.

## S2.3 Identification of positional disorder

Positional disorder is identified independently from substitutional and vacancy disorder. For each atomic site, a representative radius is assigned from oxidation-state-dependent ionic radii based on Shannon's table.[47] An empirical radius is used as instead when no oxidation state is available in the CIF. For two sites $i$ and $j$, positional disorder is assigned when the inter-site distance $d_{ij}$ satisfies:

$$d_{ij} < \max\left(1.0\ \text{Å}, 0.5(r_i + r_j)\right)$$

Where $r_i$ and $r_j$ are the representative radii of sites $i$ and $j$, respectively. The lower bound of 1.0 Å serves as a hard cutoff for clearly unphysical site-overlap, while the factor $0.5(r_i + r_j)$ provides a chemistry-aware distance criterion that scales with the radii of the two sites. The use of the 0.5 factor follows the literature criterion and was shown to suppress spurious site overlaps in ordered structures while retaining sensitivity to genuinely overlapping sites.[35]

In practice, the distances are evaluated using a periodic neighbor-list search rather than exhaustive pairwise comparison. Neighboring pairs are collected with Pymatgen's periodic neighbors searching method using a global cutoff equal to the largest possible pairwise intersection threshold in the structure. Each returned neighbor pair is then tested against its own site-specific threshold $d_{ij}$. This procedure accounts for periodic boundary conditions while remaining substantially more efficient than brute-force comparison over all site pairs and periodic images. This procedure can also identify anomalously overlapped ordered sites in general. The distinction is that such overlaps are interpreted as positional disorder when the combined occupancy of sites does not exceed 1.0.

### S3. Standardization of Wyckoff sequences

The first source of non-uniqueness symmetric description of crystal structures comes from alternative Hall settings. Hall symbols specify explicit settings of a space-group type, including the symmetry-generator choice and the corresponding crystal unique axis and cell origin convention. Consequently, the same Hermann-Mauguin space group can yield multiple Hall-equivalent Wyckoff descriptions. For example, space group 12 *C2/m* may be written as Hall 63 (−C 2y) or Hall 65 (−I 2y), which keeps the same crystal unique axis but changes the centering of crystal. In the corresponding Wyckoff-letter mapping, these two settings exchanges Wyckoff letter c and d while leaving the remaining unchanged. Therefore, Hall normalization removes this ambiguity by expressing all Hall-equivalent descriptions in one default Hall reference.

The second source of non-uniqueness remains even after the Hall setting has been fixed. A transformed Wyckoff-position (transformed WP) description is not another Hall setting, but an equivalent permutation of the Wyckoff-position table induced by a coset representative acting on the Wyckoff-position representatives. These transformations arise from the affine or Euclidean normalizer of the space group and can permute symmetry-equivalent Wyckoff positions without changing the space-group type itself. For example, in space group 12, the coset representative $(x, y, z + 0.5)$ transforms the standard listing "a b c d e f g h i j" into "c d a b f e h g i j". Thus, the same occupied orbit pattern may still appear with a different Wyckoff sequence unless these normalizer-induced transformed-WP descriptions are also canonicalized.

The two sources of non-uniqueness described above are removed in two sequential stages. Hall-setting standardization is applied first to eliminate setting-dependent relabeling. Transformed-WP standardization is then applied to remove the remaining ambiguity. Together, these two stages can be viewed as selecting a canonical representative from a set of symmetry-equivalent decorated Wyckoff sequences. To preserve the mapping between Wyckoff sites and occupied species, the object being standardized in SWORD is an elementally decorated Wyckoff sequence $S$ rather than an undecorated string of letters:

$$S = ((w_1, e_1), (w_2, e_2), \dots, (w_n, e_n))$$

where $w_n$ denotes the Wyckoff letter and $e_n$ denotes the element- or disorder-resolved occupation attached to that orbit. Standardization operates on the Wyckoff letters $w_n$, while preserving the associated chemical information $e_n$. For a given space group, the standardization is performed by enumerating all Hall-setting or Transformed WP allowed permutations of the Wyckoff sequences, from which a standardized representative will be selected. The standardized representation is defined as:

$$S^* = \arg\min_{\pi \in \Pi_{sg}} I(\pi(S))$$

Where $\Pi_{sg}$ denotes the set of all enumerated Wyckoff-letter relabelings, and $I$ maps each relabeled sequence to reference index tuple for deterministic comparison, and $arg\ min$ selects the candidate $S^*$ whose key is minimal among all equivalent relabelings. In this way, standardization is formulated as choosing one unique representative from all equivalent Wyckoff-sequence descriptions of the same occupied-orbit pattern.

In practice, this ordering is implemented by mapping Wyckoff letters to a reference index sequence and selecting the candidate with the smallest index tuple. When multiple candidates share the same leading indices, lexicographic ordering is used to break ties. This procedure ensures that all symmetry-equivalent Wyckoff sequences are reduced to a unique canonical form, independent of the original crystallographic representation.

### S4. Anomalous ICSD records identified during SWORD processing

During prescreening and labeling, a small subset of ICSD entries showed malformed or internally inconsistent crystallographic information and was therefore flagged as anomalous entries. The main anomaly classes include: (i) non-standard or placeholder species labels, such as M, X, or L; (ii) malformed CIF fields, such as corrupted fractional coordinates or invalid Wyckoff symbols; (iii) non-physical occupancies, including occupancies outside the allowed range or orbit-level occupancy sums exceeding the adopted occupancy tolerance; (iv) Wyckoff-position inconsistencies, such as mismatch between symmetry-generated multiplicity and the reported multiplicity, or sites that generate the same orbit but are reported with different representative coordinates; (v) redundantly repeated same-element, same-valence occupations on one disordered site, which are merged automatically and recorded separately as invalid disorder descriptions; and (vi) exceptionally short interatomic separations beyond the normal range of physically meaningful site overlap, which are recorded separately as too-close-site anomalies rather than treated as standard positional disorder. A small number of entries also triggered failures in the positional-disorder analysis workflow, typically due to undefined radii, occupancy inconsistencies, or non-standard CIF syntax. These flagged records were retained as a separate anomaly set for manual inspection and optional removal during ICSD curation.